\documentclass[preprint,aps]{revtex4}
\begin{document}
\title{All-optical Control of the Propagation of Intense Laser Light in Condensed Media}
\author{A. K. Dharmadhikari, K. Alti, J. A. Dharmadhikari, and D. Mathur}
\affiliation{Tata Institute of Fundamental Research, 1 Homi Bhabha Road, Mumbai 400 005, India.}
\date{\today}
\begin{abstract}
We experimentally demonstrate optical control of filamentation that occurs during propagation of intense, ultrashort laser pulses through crystals like barium fluoride and sapphire. Control is exercised by rotating the plane of polarization of the incident laser radiation and is demonstrated by directly visualizing filamentation in the bulk via six-photon absorption-induced fluorescence and, concomitantly, by probing the spectral and spatial properties of white light that is generated.   
\end{abstract}
\maketitle

Filamentation is a spectacular nonlinear process that occurs when ultrashort pulses of intense light propagate through optical media. Its importance stems from the fact that optical energy is precisely localized within a filament. The ability to localize optical energy in bulk media opens tantalizing possibilities in areas like micromachining and laser surgery. Filamentation is also important in the macroscopic domain: in air it has been used to control natural processes like lightening \cite{Kasparian} and its study has been shown to be of utility in remote sensing \cite{Diels} and in light detection and ranging \cite{Vidal}. An important ramification of the light-matter interaction in condensed media is that the localization of optical energy may make possible localized modification of material properties like refractive index. This would open up diverse applications, like three-dimensional material modification and fabrication {\it within} bulk materials without surface damage; the feasibility of ``optically fabricating" photonic devices like resonators, beam-shapers, interferometers and amplifiers has already been demonstrated \cite{Schaffer}. Periodic arrays of nanostructures have been reported in materials like fused silica \cite{rb}. Light-induced microexplosions within crystals have afforded applications in emerging areas like nanophotonics and plasmonics \cite{Juodkazis}. Such microexplosions arise from the plasma that is created when a small volume of the irradiated material (typically $\sim$0.1 $\mu$m$^3$) absorbs energy from a femtosecond laser pulse of intensity $\sim$10$^{14}$ W cm$^{-2}$. The resulting pressure (which can be in the TPa range) generates shock and refraction waves which, in turn, create nanovoids \cite{Juodkazis}. 

From a fundamental viewpoint, understanding the propagation of ultrashort light pulses in bulk media is complicated as both the temporal and spatial profiles of the incident pulse are altered due to linear and nonlinear effects like group velocity dispersion (GVD), linear diffraction, self-phase modulation (SPM), self-focusing, multiphoton ionization (MPI), plasma defocusing and self steepening \cite{Alfano}. Filamentation and white light generation are two facets of propagation effects. White light generation is a manifestation of the dynamics due to which additional frequency components are added to the incident spectrum. Recent work \cite{ourwork} has shown that SPM is responsible for generating spectral broadening that is symmetric around the incident wavelength; at higher powers, other mechanisms like self-steepening and free-electron generation due to multiphoton excitation generate blue-side frequency components, resulting in asymmetric broadening. The physics governing filament formation in transparent media is yet to be fully understood. Currently, two broad modeling approaches are in vogue. One describes a filament as a soliton-like, self-guided beam wherein filamentation dynamics are governed by competing processes like the optical Kerr effect (that results in self-focusing of the incident light, thereby increasing its intensity) and multiphoton excitation (that generates electrons, creating a low-density plasma that, in turn, defocuses the beam) \cite{Lange}. The alternative approach \cite{Brodeur} visualizes filamentation in terms of optical energy being continuously absorbed and regenerated by subsequent focusing of different time slices of the incident laser pulse. 

Can some measure of control be exercised over filamentation dynamics? An answer in the affirmative would be of obvious importance, and constitutes the subject of this Letter.  Introducing a periodic mesh creates periodic amplitude modulation of the transverse beam profile that leads to deterministic multiple filaments in liquids \cite{Chin2}. Organization of regular filamentation patterns in air has been demonstrated by imposing strong field gradients or phase distortions in the input beam profile \cite{Mysyrowicz}, with a deformable mirror \cite{Jin} or using a double lens set-up \cite{double-lens}. Increasing the ellipticity of the input beam decreases the power that is required for multiple filamentation to occur \cite{Gaeta}. For more than 30 years it has been believed that filamentation is initiated by random noise that breaks the axial symmetry of the incident beam \cite{Bespalov}. Recent work has demonstrated that noise, by itself, does not play a significant role in the filamentation dynamics \cite{Fibich} and that deterministic vectorial effects, like polarization, may be of importance. We report here polarization-dependent propagation of ultrashort laser pulses in crystals like BaF$_2$ and sapphire. We have monitored the filamentation dynamics by imaging the fluorescence signal obtained upon six-photon absorption of incident laser light in the case of barium fluoride crystals. Such visualization yields direct evidence that polarization enables us to readily exercise spatial control of filamentation within the crystal. Polarization control is also demonstrated by concomitantly monitoring the spatial and spectral properties of white light that is generated. 

We used a Ti-sapphire laser yielding 36 fs pulses (820 nm wavelength) at 1 kHz repetition rate. We used incident powers of 10-300 $\mu$J, corresponding to 10-1000 times the critical power for self-focusing. A $\lambda$/2 plate controlled the polarization angle of the light incident on each crystal. A 30 cm lens was used for loose focusing of the incident light. Crystal lengths were 3 mm for sapphire and 15 mm for BaF$_2$. The white light produced was imaged by a CCD camera and characterized by a fiber-coupled spectrometer (spanning the 400-1100 nm range). Filamentation was imaged using a digital camera and was analyzed using ImageJ software. Our crystals were randomly oriented and we rotated each around the light propagation axis so as to maximize the generation of white light. This defined the axis around which our $\lambda$/2 plate was rotated.   

The BaF$_2$ crystal affords the possibility of directly visualizing the self-focusing/defocusing events in the filamentation process. This is due to six-photon absorption-induced emission (see the blue color filaments in Fig. 1). We recall that the band gap of BaF$_2$ is 9.1 eV and the incident photon energy is ~1.5 eV. The images shown in Fig. 1 vividly demonstrate that the localization of laser energy within the crystal is amenable to ready spatial control: we spatially translate the filaments within the crystal by rotating the polarization vector, keeping the incident power constant. We quantify the polarization-controlled spatial motion in terms of z$_{of}$, the distance from the entrance face of the crystal beyond which we observe the 6PA-induced emission; this is a robust indicator of the the filament-start position. z$_{of}$ has contributions from physical focusing by the external lens as well as from self-focusing due to n$_2$, the intensity-dependent nonlinear refractive index of BaF$_2$. We fixed the position of the external lens and rotated the $\lambda$/2 plate. For data depicted in Fig. 1, the distance between the lens and crystal was 27 cm and the geometrical focus of the lens was located outside the crystal. With the $\lambda$/2 plate at 0$^o$, z$_{of}$=3.5 mm (vertical white line in Fig. 1). As the incident polarization angle rotates there is gradual spatial translation of the filament-start position. At 45$^o$ the filament-start position lies well within the crystal, with z$_{of}\sim$5.9 mm. The value of z$_{of}$ reverts to its original 3.5 mm when the $\lambda$/2 plate is at 90$^o$. The overall change in z$_{of}$ is significantly large, as much as 16\% of the crystal length. 

Can the physical shift of z$_{of}$ be rationalized within the context of existing knowledge? To probe this, we take recourse to methods used in related numerical studies of light-induced damage in bulk media wherein z$_{of}$ has been related to experimental parameters \cite{Junnarkar} like focal length, size of the beam, the distance between the lens and the entrance face of the crystal, the incident laser power and P$_{cr}$, the critical power for self focusing which, for isotropic materials, is (3.77$\lambda^2$)/(8 $\pi$n$_o$n$_2$). We note that P$_{cr}$ depends inversely on n$_2$ which contributes to the net refractive index of a material: n = n$_o$ + n$_2|E|^2$, where n$_0$ is the linear refractive index. n$_2$ is further related to the material's third-order susceptibility tensor: n$_2$ = (12$\pi$/n$_o$)$\chi^{(3)}$. In case of cubic crystals like BaF$_2$ with space-group symmetry $m3m$, the effective value of $\chi^{(3)}$ is \cite{Payne}:
\begin{equation}
\chi^{(3)}(\theta) = {3 \chi^{(3)}_{1122} + (\chi^{(3)}_{1111} - 3 \chi^{(3)}_{1122})}{{[{\rm cos^2}(\theta) + 1]}\over{2}},
\end{equation}
where $\theta$ is the incident polarization angle. As $\theta$ changes, the effective value of $\chi^{(3)}$ also changes and this, in turn, alters n$_2$. The n$_2$ value along the $e$-direction depends on the propagation direction while along the $o$-direction is independent of the direction of propagation. Thus, P$_{cr}$ is direction dependent. In order to account for the observed shifts in z$_{of}$, for fixed incident power, the value of P$_{cr}$ would be required to change sixfold! This is clearly an unrealistic proposition. We calculate that the change in n$_2$ for BaF$_2$ is only 0.43$\times$10$^{-16}$ cm$^2$ W as $\theta$ changes by 45$^o$, resulting in only a 130\% change in P$_{cr}$. Other factors are clearly at play and need to be identified and explored. 

Nonlinear absorption effects will not affect the beam power at the onset of filamentation because these will set in only after sufficient intensity has been attained. Indeed, the beam power may actually decrease with propagation due to GVD-induced pulse broadening. However, this effect does not significantly broaden the pulse as z$_{of}$ changes from 3.5 mm to 5.9 mm. Other factors that we have considered include refractive index modification, color center generation, and induced birefringence. The loose focusing employed in our experiments rules out the possibility of modifying refractive index and generating color centers; melting \cite{melting} is also ruled out. However, we note that loose focusing might well give rise to cross-polarized wave generation wherein a single input wave is partly converted to a wave that is polarized perpendicular to the input plane polarization due to degenerate four wave mixing. Such a process has recently been shown to occur with high efficiency in BaF$_2$ \cite{Jullien} and may lead to variation in the critical power for self-focusing, subsequently changing the onset of filamentation that is observed in Fig. 1.

For uniaxial crystals like sapphire, the relation for effective $\chi^{(3)}$ is considerably more complicated as there is now a direction-dependence for $e$-rays but not for $o$-rays \cite{Chi}. The value of P$_{cr}$ will vary accordingly. In sapphire, where filamentation was not visualized due to the absence of six-photon-absorption-induced emission, we use white light spectra as the diagnostic. White light spectra from sapphire (see Fig. 2a) have two components: one due to SPM that arises from the Kerr nonlinearity (symmetric about the incident wavelength) and an asymmetric component that arises from MPI-induced free electron generation. The latter contribution is on the blue side of the spectrum and in all our measurements we saw a polarization-dependent change in this blue-side component. The Kerr nonlinearity that we observe depends, as noted above, on n$_2$ and $\chi^{(3)}$. The latter is a tensor with 81 components. Depending on crystal type and symmetry the tensor may have many non-zero components contributing to the effective $\chi^{(3)}$ which will result in change in the value of n$_2$. Other experiments on isotropic materials have shown that at high intensities there are deviations from isotropy due to induced birefringence \cite{George}: the induced birefringence makes any isotropic material behave like a uniaxial crystal.  

We also made simultaneous measurements of polarization-dependent white light spectra BaF$_2$ (Fig. 2b). The 820 nm light was blocked in this case; only the white light was recorded. Symmetric and asymmetric components are clearly visible. Recent studies have explored variations in the efficiency of white light production and of its spectral content as a function of incident laser energy \cite{ourwork} but, to our knowledge, studies of how the overall shape of the spectra change as a function of polarization angle have not been reported. Data depicted in Fig. 2c show that there is a major polarization-dependent change in the white light spectral distribution, specifically a reduction in the spectral width as the incident laser polarization nears 45$^o$. As the blue part of the spectrum is predominantly due to plasma effects (free electron generation by MPI \cite{ourwork}), the reduction in width sensitively maps the lessening of effective laser intensity within the medium which is driven by alterations in self-focusing conditions that are a consequence of changes in local refractive index. Over and above measurement of spectra, it is also instructive to directly visualize the white light pattern. Results are shown in Fig. 3 for BaF$_2$ irradiated at different polarization angles that demonstrate another manifestation of filament translation. When the filament-start is close to the crystal entrance the white light spot is large. As the half-wave plate is rotated, the filament-start position moves inside the crystal, reducing the white spot in the far field. We observed more than a two-fold reduction in the white light spot size when the polarization angle was rotated by 45$^o$. At 0$^o$ the central white light zone is surrounded by concentric rings called conical emission; the rings depend on SPM in the radial direction. As the polarization angle is changed the ring contrast degrades and the spatial extent of the conical emission is reduced: in the radial direction only those frequencies that satisfy the phase matching condition form concentric rings. The diameter of the central white light spot is polarization dependent, showing a minimum at 45$^o$. Similar results were observed for sapphire. We note that self-induced polarization in isotropic media has also been recently reported by Midorikawa and coworkers \cite{midorikawa}.

To place our polarization-dependent control results in perspective, we note that it has been possible to achieve a measure of control on filament morphology in air by forcing aberrations in the incident light beam and by means of diaphragms \cite{Mysyrowicz,Dubietis}. Mysyrowicz and coworkers \cite{Mysyrowicz} have shown that by varying the laser power the self-focusing distance in air can be altered. We also made measurements at various incident powers; Figure 4 shows how the onset of filamentation depends on laser power. At relatively low power we see a single filament whose start position lies well inside the crystal (4.6 mm from the front face). As the power increases, the filament-start position (z$_{of}$) moves nearer the entrance face of the crystal. The most noteworthy feature is that in order to effect a change in filament-start position of $\sim$3 mm, as much as a 5-fold increase in power is required. The superiority of our simple polarization method of controlling the position of the filament-start position is clearly demonstrated. Our findings have diverse ramifications in applied areas and in the basic sciences, particularly from the viewpoint of material modification in the bulk. 

Useful discussions with See-Leang Chin and Andr\'e Bandrauk are acknowledged. The Homi Bhabha Fellowship Council is thanked for financially supporting one of us (JAD).

\newpage

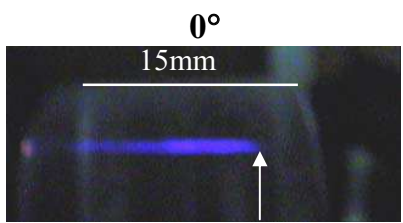
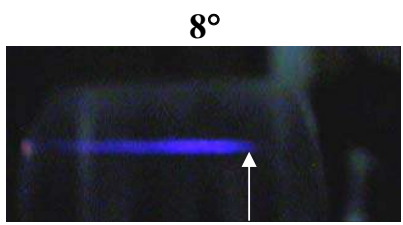
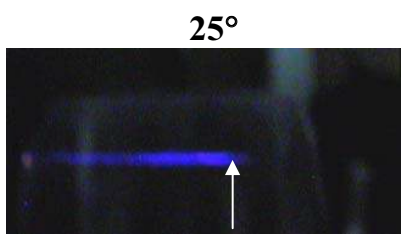
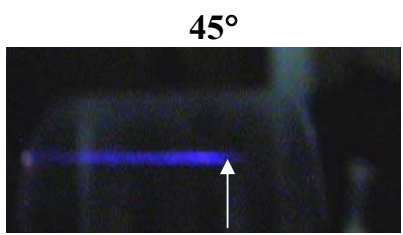
\begin{figure}
\caption{Polarization-dependent control of the spatial location at which filamentation commences in a BaF$_2$ crystal. The horizontal white line marked 15 mm indicates the spatial extent of the crystal. Vertical white lines are a guide to the shift in filament-start position with polarization angle.}
\end{figure}

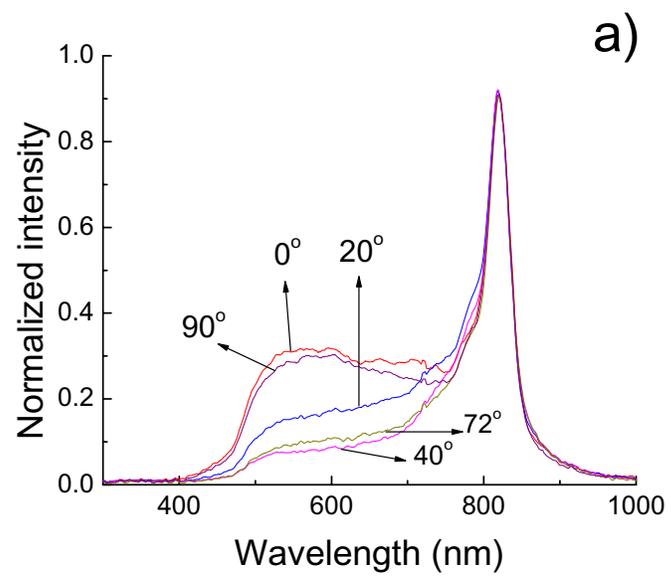
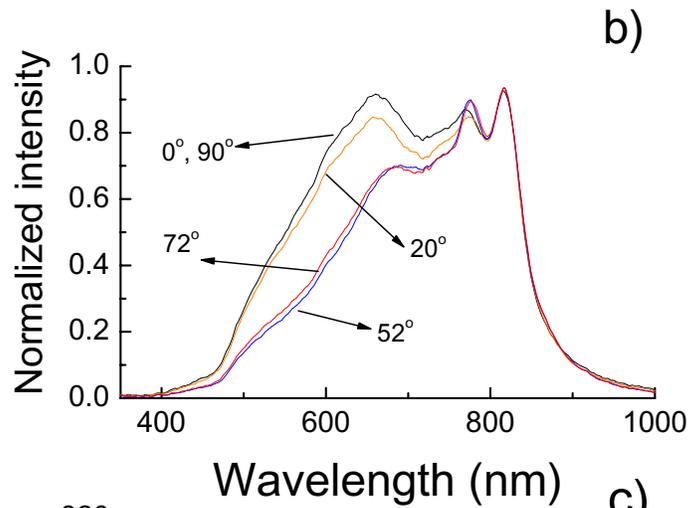
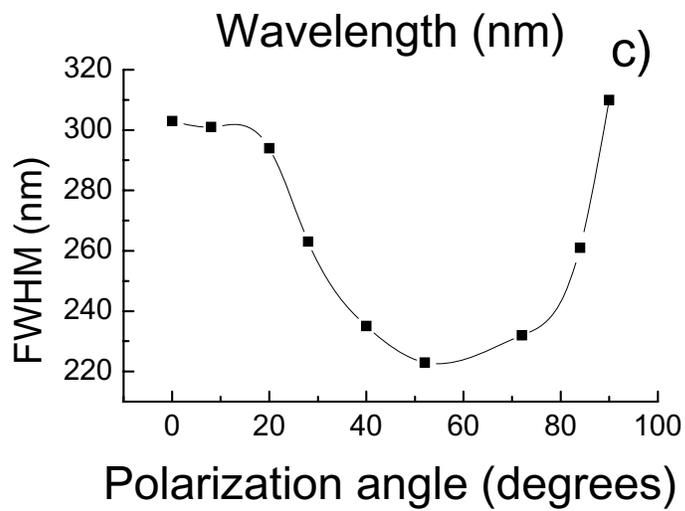
\begin{figure}
\caption{Polarization-dependent white light spectra obtained upon irradiation of a) a 3 mm long sapphire crystal (50 $\mu$J) and b) a 15 mm long BaF$_2$ crystal (140 $\mu$J).  c) Polarization-dependent full width at half maximum (FWHM) of white light spectra from BaF$_2$.}
\end{figure}

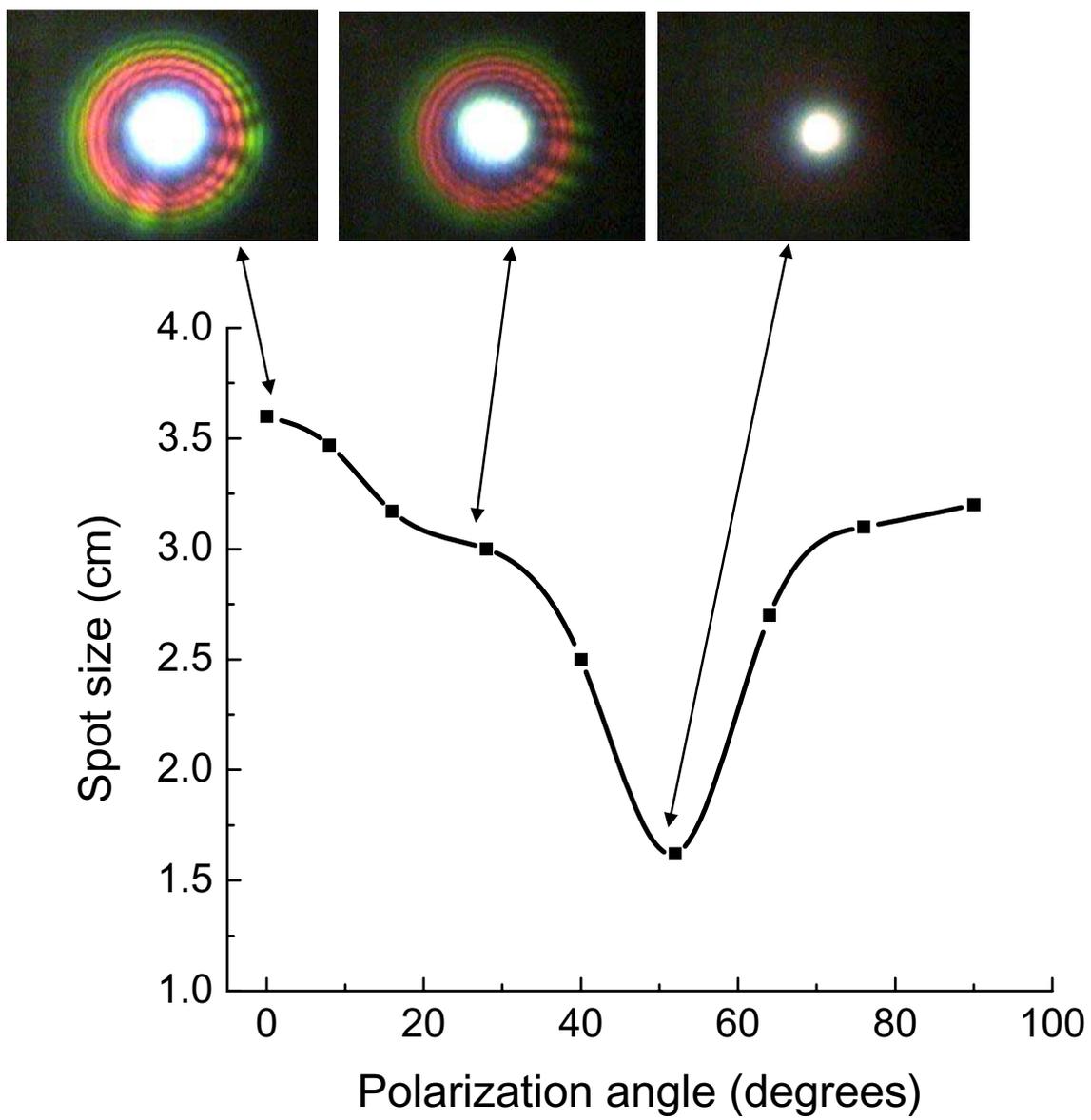
\begin{figure}
\caption{Direct visualization of the size of the white light spot in BaF$_2$ as a function of polarization angle (16 $\mu$J). Note also the angle-dependence exhibited by the conical emission.}
\end{figure}

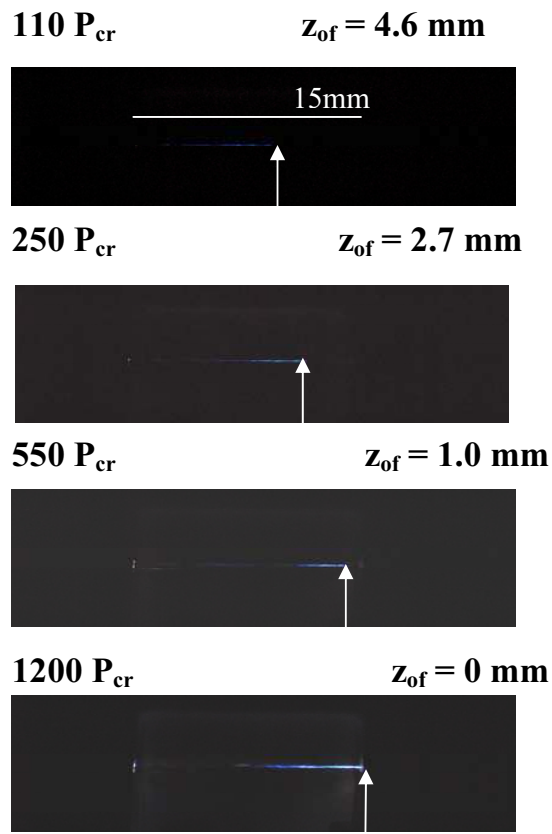
\begin{figure}
\caption{Variation of the onset of filamentation in BaF$_2$ with incident laser power. The spatial control is expressed in terms of z$_{of}$, the distance from the entrance face of the crystal to the filament-start position.}
\end{figure}

\end{document}